%% file: main.tex
\documentclass[authoryear,preprint,review,12pt]{elsarticle}

\usepackage{amssymb}
\usepackage{amsthm}
\input{preamble}

\journal{}

\begin{document}

\begin{frontmatter}

%% Title, authors and addresses

%% use the tnoteref command within \title for footnotes;
%% use the tnotetext command for theassociated footnote;
%% use the fnref command within \author or \affiliation for footnotes;
%% use the fntext command for theassociated footnote;
%% use the corref command within \author for corresponding author footnotes;
%% use the cortext command for theassociated footnote;
%% use the ead command for the email address,
%% and the form \ead[url] for the home page:
%% \title{Title\tnoteref{label1}}
%% \tnotetext[label1]{}
%% \author{Name\corref{cor1}\fnref{label2}}
%% \ead{email address}
%% \ead[url]{home page}
%% \fntext[label2]{}
%% \cortext[cor1]{}
%% \affiliation{organization={},
%%            addressline={}, 
%%            city={},
%%            postcode={}, 
%%            state={},
%%            country={}}
%% \fntext[label3]{}

\title{Using Staged Tree Models for Health Data: Investigating Invasive Fungal Infections by Aspergillus and Other Filamentous Fungi}

%% use optional labels to link authors explicitly to addresses:
%% \author[label1,label2]{}
%% \affiliation[label1]{organization={},
%%             addressline={},
%%             city={},
%%             postcode={},
%%             state={},
%%             country={}}
%%
%% \affiliation[label2]{organization={},
%%             addressline={},
%%             city={},
%%             postcode={},
%%             state={},
%%             country={}}

\author[1]{Maria Teresa Filigheddu} 
\author[IE]{Manuele Leonelli}
\author[UV]{Gherardo Varando} 
\author[4]{Miguel Ángel Gómez-Bermejo}
\author[4]{Sofía Ventura-Díaz}
\author[4]{Luis Gorospe}
\author[1,5]{Jesús Fortún}

\affiliation[1]{organization={Infectious Diseases Department, Hospital Ramón y Cajal, IRYCIS (Instituto
Ramón y Cajal de Investigación Sanitaria); Universidad de Alcalá}, city =  {Madrid}, country =  {Spain}}
\affiliation[IE]{organization={School of Science and Technology, IE Univeristy},
city = {Madrid},%Department and Organization
            country={Spain}}
\affiliation[UV]{organization={Image Processing Laboratory (IPL), Universitat de València},%Department and Organization
            city={Valencia},
            country={Spain}}
\affiliation[4]{organization = {Radiology Department, Hospital Universitario Ramón y Cajal}, city = {Madrid}, country = {Spain}}
\affiliation[5]{organization = {Microbiology Department, Hospital Universitario Ramón y Cajal}, city = {Madrid}, country = {Spain}}

\begin{abstract}
Machine learning models are increasingly used in the medical domain to study the association between risk factors and diseases to support practitioners in predicting health outcomes. In this paper, we showcase the use of machine-learned staged tree models for investigating complex asymmetric dependence structures in health data. Staged trees are a specific class of generative, probabilistic graphical models that formally model asymmetric conditional independence and non-regular sample spaces. An investigation of the risk factors in invasive fungal infections demonstrates the insights staged trees provide to support medical decision-making.
\end{abstract}

%%Graphical abstract
%\begin{graphicalabstract}
%\includegraphics{grabs}
%\end{graphicalabstract}

%%Research highlights
%\begin{highlights}
%\item Research highlight 1
%\item Research highlight 2
%\end{highlights}

\begin{keyword}
%% keywords here, in the form: keyword \sep keyword
Diagnostic criteria \sep Invasive aspergillosis\sep Machine learning \sep Probabilistic graphical models \sep Staged trees
%% PACS codes here, in the form: \PACS code \sep code

%% MSC codes here, in the form: \MSC code \sep code
%% or \MSC[2008] code \sep code (2000 is the default)

\end{keyword}

\end{frontmatter}

%% \linenumbers

%% main text
\section{Introduction}
Risk factors increase an individual's likelihood of negative outcomes for a  particular disease or condition. 
They can range from demographic factors such as age, gender, and ethnicity to behavioral factors such as physical inactivity, smoking, and unhealthy diet. Environmental factors, such as air and water pollution, exposure to radiation, and access to healthcare, also play a role in determining an individual's risk for certain diseases. The study of risk factors and their relationship to medical outcomes is critical in medicine as it can provide insights into the causes of diseases, help identify populations at high risk, and inform the development of prevention and treatment strategies.

Machine learning techniques have emerged as powerful tools for modeling and predicting the relationships between risk factors and medical outcomes. These techniques can analyze large and complex datasets, identify patterns and relationships that are not immediately apparent, and provide predictions with high accuracy. For example, machine learning algorithms have been used to predict the risk of cardiovascular disease in individuals based on electronic health records~\citep{ahsan2022machine} and to identify the risk of developing diabetes in a large population-based cohort~\citep{liao2022development}. As a result, machine learning has become an essential tool for medical research, helping health practitioners, researchers, and policymakers to understand better the impact of risk factors on health outcomes and to develop targeted interventions to prevent or mitigate these risks.

Machine learning techniques have emerged as powerful tools for modeling and predicting the relationships between risk factors and medical outcomes. Recent trends in applied machine learning focused on developing highly complex and black-box models that, while providing high levels of prediction accuracy, lack interpretability and intuitiveness \citep[e.g.][]{alam2019random,painuli2022recent,sun2022deep}. However, there is an increasing awareness of the criticality of developing AI systems that can provide clear and interpretable explanations for their decision-making process. Explainable AI (XAI) can help address the trust and interpretability issues associated with black-box models and increase users' trust and adoption \citep{amiri2021peeking,lime:kdd16}.

Probabilistic graphical models \citep{koller2009probabilistic} are generative machine learning models that visually represent the overall dependence structure using graphs. They do not simply model the conditional distribution of the output of interest given the available risk factors, as in discriminative models, but the overall probability distribution. They thus provide an intuitive platform to perform inferential and independence queries, sensitivity analyses, and risk factors' rankings: all critical activities in applied machine learning modeling \citep{kjaerulff2008bayesian}.

This article showcases the use in medical research of a relatively new class of graphical models called \textit{staged trees} \citep{collazo2018chain,smith2008conditional}. They are probability trees whose inner vertices are colored to embed conditional independence information formally. Recent critical advances have made them a viable, efficient, and highly informative alternative to competitor models in health applications. In particular: (i) a comprehensive software package now implements staged trees for the e of practitioners in any area of science \citep{Carli2022}; (ii) faster and more flexible algorithms to learn staged trees from data have been recently developed, scaling them up to dozen of variables \citep[e.g.][]{leonelli2022highly,shenvi2022beyond}; (iii) novel visualization frameworks allow for an intuitive depiction of the underlying dependence structure between the variables \citep{varando2021staged}; (iv) recent theoretical advances have formalized the use of staged trees for classification problems \citep{carli2020new} and causal reasoning and discovery \citep{leonelli2023context,thwaites2013causal}.

We showcase the use of staged trees for health data by studying the risk factors associated with invasive fungal infections. Invasive fungal infections by aspergillus and other filamentous fungi (henceforth called AFF-IFI) have become a critical public health problem in recent decades~\citep{bongomin2017global}, with an increase in filamentous fungal infections mainly due to Aspergillus spp, and to a lesser extent, Mucor spp and Fusarium spp \citep{lass2009changing}. AFF-IFIs are significantly related to impaired host immune response, having generally affected patients with hematological malignancies, those undergoing hematopoietic stem cell transplantation, or solid organ transplantation under immunosuppressive treatment \citep{de2008revised}.

The most widely used diagnostic criteria \citep{de2008revised,donnelly2020revision} include host factors (mainly immunocompromised conditions), radiological findings (highly specific but also very strict and, as such, are unsuitable for less immunocompromised patients), and microbiological tests. They have been validated on oncohematological patients with a high level of immunosuppression only. However, an increasing number of patients with impaired immunity from other causes have been observed, with various systemic or bronchopulmonary pathologies \citep[among others, critically ill patients admitted to the intensive care unit (ICU) with severe influenza or other viral infections, repeated use of corticosteroids, poorly controlled diabetes, etc.;][]{latge2019aspergillus} with the consequent progressive increase in AFF-IFI in non-oncohematological patients and increased mortality, possibly due to lack of clinical suspicion \citep{menzin2009mortality}.

Diagnosing these patients is not always easy because the AFF-IFI of non-oncohematological patients differs considerably in the degree of immunosuppression and underlying pathologies. Although they have as a common denominator the development of respiratory forms more frequently with bronchopulmonary expression, they are not always identifiable, unlike oncohematological patients who more frequently favor more evident forms of hematogenous expression to diagnose \citep{de2008revised,donnelly2020revision}.

The progressive increase in AFF-IFI in non-oncohematological patients with various systemic or bronchopulmonary pathologies has been associated, contrary to expectations, with an increase in mortality, possibly due to a lack of clinical suspicion \citep{menzin2009mortality}. In particular, due to its prevalence and expensive treatment, it has become the most expensive fungal disease in the hospital setting \citep{benedict2019estimation}.

The emergent threat of invasive fungal diseases is driven by antifungal resistance and limited global access to diagnostic tools and treatments \citep{brown2012hidden, whoFungal}.
This menace has vast implications for public health worldwide. It usually gives on to more extended hospital stays and treatments, including the need for expensive antifungal medicines, which are often unavailable in developing countries~\citep{van2011clinical, whoFungal},
Even if fungal infections have been recently recognized as a growing threat to human health worldwide, 
their study and clinical monitoring receive little resources at a global level~\citep{bongomin2017global}.
This fact hinders our understanding of the problem and makes it impossible to understand its exact burden on public health \citep{whoFungal}.

Machine learning techniques for risk prediction and factor identification have only recently started to be used in AFF-IFIs \citep{li2022machine,mayer2022machine,yan2022machine,yuan2021using}. Only \citet{potter2019combat} developed a decision support system based on probabilistic graphical models for combat-related AFF-IFI patients.  With the present study, we contribute to developing robust and interpretable machine-learning approaches to understanding AFF-IFI diseases.

The two case studies below highlight the critical need to devise new diagnostic criteria, more widely applicable than the gold standard~\citep{de2008revised,donnelly2020revision}, including the overlooked non-conventional group and broncoinvasive patterns, instead of angioinvasive ones only.

\subsection{Health Applications}

Data-driven algorithms for discovering the genetic causes of various diseases are most commonly based on DAGs \citep[e.g.][]{corander2022causal,foraita2020causal,pingault2022causal}. Applications of such causal discovery algorithms for understanding clinical risk factors have also recently started to appear \citep{tennant2021use,VELIKOVA201459}. Conversely, the use of BNs as a decision-support tool for practitioners and as a platform for the study of risk factors is not as widespread \citep{kyrimi2020bayesian, kyrimi2021comprehensive,kyrimi2021bayesian, mclachlan2020bayesian}, although recently they have been more frequently used \citep[e.g.][]{song2023using,tian2023bayesian,van2022feasibility}.

Tree-based machine learning algorithms are standard in health, for instance, decision trees \citep{doupe2019machine,kilic2020artificial}. However, these are not probability trees formally and highly differ from staged trees. 
Simple probability (or frequency) trees are vastly used in 
health and medical literature~\citep{ LEFRANCQ2021} 
and are helpful to practitioners in computing probability queries~\citep{detsky1997primer, binder2018visualizing}.
Staged trees have been used in four medical applications in the past to investigate: type I diabetes \citep{keeble2017learning}, the effect of social and economic factors on kids' health \citep{barclay2013refining}, covid-19 trajectories \citep{leonelli2023context}, and data missingness patterns in health data \citep{barclay}. 

\section{Materials and Methods}

\subsection{Data}

Retrospectively included in the study were all non-oncohaematological patients diagnosed with proven, probable, or possible pulmonary AFF-IFI according to different diagnostic scales during 1998-2021 in 3 hospitals in the
Community of Madrid, Spain (Hospital Ramón y Cajal, Hospital Doce de Octubre, Hospital Universitario Fundación Alcorcón) 
and two hospitals in the metropolitan city of Cagliari, Autonomous Region of Sardinia (Ospedale Santissima Trinitá di Dio and Ospedale Azienda Sanitaria G. Brotzu). 
The radiological findings were reviewed and described by three radiologists blinded to patients' characteristics.

The chest CT findings and radiological pattern (RP, categorized as angioinvasive or broncoinvasive)) of 146 patients\footnote{The small sample size is due to the limited widespread of AFF-IFI. Related studies include comparable  patients" samples if not smaller~\citep{liu2020airway,park2010clinical}.} with pulmonary AFF-IFI and its prognostic value in non-oncohaematological patients divided into three groups (GR) according to the degree of immunosuppression have been assessed. The first group includes patients with neutropenia not related to haematological diseases (neutropenic group); the second includes patients who do not have neutropenia and have at least one of the following: solid organ transplant and/or tumour, inflammatory/autoimmune diseases, congenital or acquired immunodeficiency, or use of corticosteroids (conventional group); the third includes patients who, in the absence of neutropenia or another ``conventional" immunosuppression factor, present alterations in innate and/or adaptive immunity described in the literature as related to specific populations at risk of AFF-IFI (non-conventional group). Despite the known relationship between this group and AFF-IFI mortality, their risk factors are not included in the most widely used diagnostic scales. Therefore they are often not diagnosed with AFF-IFI. 

Moreover, we consider the patient trajectory as recorded in the hospital: if entered the intensive care unit (ICU), if intubation is performed (IN) and finally the 
survival outcome (DTH).

In a second case study, we consider additional risk factors for AFF-IFI that are known to be individually associated with an increase in mortality~\citep{gioia2021invasive,park2010clinical}. These are reported in Table~\ref{table:variables} together with 
previous variables.

\begin{table}
  \caption{Variables considered for the population under study, acronyms, and variables' levels. Sample distribution in each level is shown in parentheses.\label{table:variables}}
  
  \scalebox{0.6}{
   \renewcommand{\arraystretch}{1.2}
\begin{tabular}{c|c|c}
\hline
Variable & Acronym & Levels\\
\hline
\textbf{\cellcolor{gray!6}{GROUP}} & \cellcolor{gray!6}{GR}& \cellcolor{gray!6}{Neutropenic (9), Conventional (105), Non-conventional (32)}\\
\hline
\textbf{RADIOLOGICAL PATTERN} & RP & Angioinvasive (80), Broncoinvasive (66)\\
\hline
\textbf{\cellcolor{gray!6}{ICU}} & \cellcolor{gray!6}{ICU}& \cellcolor{gray!6}{No (74), Yes (72)}\\
\hline
\textbf{INTUBATION} & INT & No (80), Yes (66)\\
\hline
\textbf{\cellcolor{gray!6}{DEATH}} & \cellcolor{gray!6}{DTH}& \cellcolor{gray!6}{No (91), Yes (55)}\\
\hline
\textbf{IMMUNOTHERAPY} & IM & No (54), Yes (77) \\
\hline
\textbf{\cellcolor{gray!6}{SYSTEMIC CORTICOIDS}} & \cellcolor{gray!6}{SC}& \cellcolor{gray!6}{No (91), Yes (40)}\\
\hline
\textbf{PREVIOUS NEUTROPENIA} & PN & No (123), Yes (8)\\
\hline
\textbf{\cellcolor{gray!6}{VIRAL PNEUMONIA}} & \cellcolor{gray!6}{VP}& \cellcolor{gray!6}{No (103), Yes (28)}\\
\hline
\textbf{CMV INFECTION} & CMV & No (99), Yes (32)\\
\hline
\textbf{\cellcolor{gray!6}{DIAGNOSTIC TIME}} &\cellcolor{gray!6}{DT}& \cellcolor{gray!6}{$<16$ (54), $\geq 16$ (77)}\\
\hline
\textbf{SOLID ORGAN TRANSPLANT} & SOT & No (79), Yes (52)\\
\hline
\textbf{\cellcolor{gray!6}{MALNUTRITION}} & \cellcolor{gray!6}{MN}& \cellcolor{gray!6}{No (46), Yes (85)}\\
\hline
\end{tabular}
}
\end{table}

\subsection{Logistic Regression}

Simple univariate logistic regressions for variables ICU, INT, and DTH are fitted to data, using as predictor GR, RP, and the preceding variables in the trajectory (ICU $\rightarrow$ INT $\rightarrow$ DTH).  

To study the relationship between the predictors and their effect on the probability of death, we fitted a group LASSO logistic regression~\citep{yang2015fast}, where the regularization parameter was chosen via a 5-fold cross-validation. The model includes as predictors also all 2-way interactions.

\subsection{Bayesian Networks}

Evolving from the path coefficients
method of~\cite{wright1934method}, 
Bayesian Network (BN) 
models~\citep{pearl1988probabilistic} have become
powerful tools in data science and 
statistics~\citep{bielza2014discrete, bielza2014bayesian}. 
A BN defines a factorization of a random vector's probability mass function (pmf) using a directed acyclic graph (DAG). More formally, let $[p]=\{1,\dots,p\}$ and $\bm{Y}=(Y_i)_{i\in[p]}$ be a random vector of interest with sample space $\mathbb{Y}=\times_{i\in[p]}\mathbb{Y}_i$. A BN defines the pmf $P(\bm{Y}=\bm{y})$, for $\bm{y}\in\mathbb{Y}$, as a product of simpler conditional pmfs as follows:
\begin{equation}
P(\bm{Y}=\bm{y}) = \prod_{i\in[p]}P(Y_i=y_i\;|\; \bm{Y}_{\Pi_i}=\bm{y}_{\Pi_i}),
\end{equation}
where ${\Pi_i}$ are the parents of $i$ in the DAG associated to the BN. Assuming variables are topologically ordered, the BN is moreover, defined by the (symmetric) conditional independence statements $Y_i\independent Y_{[i-1]}|Y_{\Pi_i}$.

The DAG associated with a BN provides an intuitive overview of the relationships between variables of interest. However, it also provides a framework to assess if any generic conditional independence holds for a specific subset of the variables via the so-called d-separation criterion \citep[e.g.][]{pearl1988probabilistic}. Furthermore, the DAG provides a framework for the efficient propagation of probabilities and evidence via algorithms that take advantage of the structure of the underlying DAG~\citep{cowell2007probabilistic}.

Although the underlying DAG can be elicited using expert judgment, it is most commonly learned from data using nuanced optimization algorithms~\citep[e.g.][]{scutari2019learns, glymour2019review}.
Moreover, BNs and DAGs are the gold standards for representing and learning causality from data, providing an intuitive framework for defining causal interventions and predicting their effects~\citep{pearl2009causality,peters2017elements,glymour2019review}.

%Bor build classifier N can be  defined for general random variables; they are extensively used,  for instance, with Gaussian distributions~\citep{geiger1994learning,koller2009probabilistic}. 

We fit a BN over the same five variables as the logistic regression (GR, RP, ICU, INT and DTH). Since BNs are generative models, they estimate a full probability distribution over all variables, formally modelling their dependence. In particular, a BN is learned using 5000 bootstrap replications of the \texttt{tabu} algorithm implemented in \texttt{bnlearn}~\citep{Scutari2010}, enforcing the causal order GR, RP, ICU, INT, and DTH. Arcs appearing in more than 50\% of the replications are then retained in the final model. 

Similarly in the second case study (Section~\ref{sec:case2}) a 
BN is fitted using the same procedure over all the variables in Table~\ref{table:variables} but INT. 

\subsection{Staged Trees}

There is an increasing awareness that the stringent assumption of \emph{symmetric} conditional independence of DAGs may be too restrictive in applications~\citep{mokhtarian2022causal,pensar2016role,tikka2019identifying}. The most common non-symmetric conditional independence is \emph{context-specific}~\citep{Boutilier1996}: the independence between two variables holds only for  specific values (called this context) of conditioning variables: e.g. $Y_i\independent Y_j | Y_k = y_k$ for a specific $y_k\in \mathbb{Y}_k$. 
More flexible and generic types of independence statements have been defined, namely partial and local~\citep{pensar2016role}.

Although models accounting for non-symmetric independence have been defined, staged trees~\citep{collazo2018chain,smith2008conditional} are the only ones 
 extensively studied and implemented in user-friendly software~\citep{Carli2022}. Since a precise definition of staged trees is beyond this paper's scope and can be found elsewhere~\citep{duarte2021representation,varando2021staged}, we introduce them next with an example.

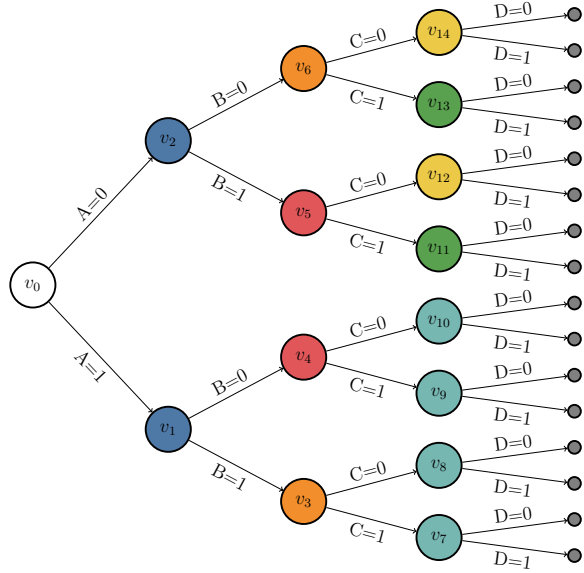
\begin{figure}
\centering
\scalebox{0.6}{
\input{trees/tree_example}
}
\caption{Example of a staged trees over four binary random variables $X_1,X_2,X_3,X_4$. \label{fig:st}}
\end{figure}

Figure~\ref{fig:st} reports a staged tree over four binary random variables $A, B, C, D$, each taking values in $\{0,1\}$. Each root-to-leaf path represents an atomic event (an assignment of all four variables), and inner vertices with corresponding emanating edges are associated with conditional probabilities. For instance, $v_1$ is associated with the conditional distribution of $B$ conditional on $A=0$. Colors of the inner vertices are interpreted as equality of conditional distributions: the blue vertices $v_1$ and $v_2$ denote that $P(B|A=0)=P(B|A=1)$, or equivalently $A\independent B$.

The flexibility of the coloring allows for asymmetric forms of independence. For instance the cyan colored vertices $v_{7}-v_{10}$ denote the context-specific independence $D\independent B, C|A=1$. 
Instead, the yellow and green vertices represent 
$D\independent B| C, A=0$.
Finally, red and orange vertices (at depth 2 from the root) are associated with so-called local independences~\citep{pensar2016role}, where no discernable patterns can be identified since they imply 
equalities $P(C|A=0, B=0)=P(C|A=1, B=1)$ and $P(C|A=0, B=1)=P(C|A=1, B=0)$.
Every BN can be represented exactly as a staged tree~\citep{varando2021staged}, while the reverse does not hold since the coloring allows for non-symmetric independence statements which have no graphical representation in a DAG. 

Although, as for BNs, staged trees could be elicited from experts, they are most often learned from data using heuristic algorithms~\citep{Carli2022, leonelli2022highly}.

%The stringent assumption of symmetric conditional independence in standard BNs, which are not flexible enough to learn complex dependence patterns, could 
%in general result in more sparse (or simpler) models. 
We employ a staged tree model to provide a flexible picture of patients' trajectories and outcomes. A priori, we fix the ordering of the variables as GR, RP, ICU, INT, and DTH to represent the actual steps of the patient's trajectory.

For the second case study, learning a staged tree over eleven binary and one ternary variable is challenging because of the exponentially growing model space size \citep[e.g.][]{duarte2021representation}. Visualizing the staged tree would also be impossible as it would include $2^{11}\cdot 3=6144$ leaves. To circumvent the visualization issue,
\citet{varando2021staged} defined the so-called \textbf{minimal DAG}: a DAG representation of the staged tree such that two variables are d-separated in the minimal DAG if and only if the coloring of the staged tree embeds the associated conditional independence. 

Because of the high flexibility of the coloring of staged trees, minimal DAGs of staged trees are usually fully connected unless some sparsity is imposed in the structural learning algorithms. Sparsity is the gold standard in Gaussian probabilistic graphical models \citep[e.g.][]{friedman2008sparse}. In the context of discrete BNs, one of the first attempts to impose sparsity was to limit the number of parents each variable can have \citep{friedman1999learning,tsamardinos2006max}. This makes sense from an applied point of view since, most often, only a limited number of variables can be expected to influence another directly. Setting a maximum number of parents is also available in the standard \texttt{bnlearn} software \citep{Scutari2010}.

Limiting the maximum number of parents further decreases the size of the possible models, speeding up structural learning algorithms. For this reason, \citet{leonelli2022highly} introduced learning algorithms for \textbf{$k$-parents} staged trees: staged trees whose minimal DAG has a maximum in-degree less or equal to $k$.

\section{Results}

\subsection{1st Case Study}

\begin{table}
  \caption{Odds ratios, confidence intervals and p-values for univariate logistic regressions to predict death. \label{table:logistic}}
  \scalebox{0.6}{
  \renewcommand{\arraystretch}{1.2}
\begin{tabular}{ccccccc}
\toprule
Response & Predictor & Term & OR & CI.Low & CI.High & P-value\\
\midrule
{\cellcolor{gray!6}{DEATH}}&{\cellcolor{gray!6}{GROUP}} & \cellcolor{gray!6}{Neutropenic (Intercept)} & \cellcolor{gray!6}{0.80} & \cellcolor{gray!6}{0.20} & \cellcolor{gray!6}{3.02} & \cellcolor{gray!6}{0.74}\\
{DEATH}&{GROUP} & Conventional & 0.71 & 0.18 & 3.01 & 0.62\\
{\cellcolor{gray!6}{DEATH}}&{\cellcolor{gray!6}{GROUP}} & \cellcolor{gray!6}{Non-Conventional} & \cellcolor{gray!6}{1.10} & \cellcolor{gray!6}{0.25} & \cellcolor{gray!6}{5.18} & \cellcolor{gray!6}{0.90}\\
{DEATH}&{RADIOLOGICAL PATTERN} & Bronco. (Intercept) & 0.51 & 0.32 & 0.80 & 0.00\\
{\cellcolor{gray!6}{DEATH}}&{\cellcolor{gray!6}{RADIOLOGICAL PATTERN}} & \cellcolor{gray!6}{Angio.} & \cellcolor{gray!6}{1.64} & \cellcolor{gray!6}{0.84} & \cellcolor{gray!6}{3.22} & \cellcolor{gray!6}{0.15}\\
{DEATH}&{ICU} & No (Intercept) & 0.37 & 0.22 & 0.61 & 0.00\\
{\cellcolor{gray!6}{DEATH}}&{\cellcolor{gray!6}{ICU}} & \cellcolor{gray!6}{Yes} & \cellcolor{gray!6}{2.85} & \cellcolor{gray!6}{1.44} & \cellcolor{gray!6}{5.77} & \cellcolor{gray!6}{0.00}\\
{DEATH}&{INTUBATION} & No (Intercept) & 0.40 & 0.24 & 0.65 & 0.00\\
{\cellcolor{gray!6}{DEATH}}&{\cellcolor{gray!6}{INTUBATION}} & \cellcolor{gray!6}{Yes} & \cellcolor{gray!6}{2.63} & \cellcolor{gray!6}{1.34} & \cellcolor{gray!6}{5.27} & \cellcolor{gray!6}{0.01}\\
\midrule
{INTUBATION}&{GROUP} & Neutropenic (Intercept) & 0.13 & 0.01 & 0.68 & 0.05\\
{\cellcolor{gray!6}{INTUBATION}}&{\cellcolor{gray!6}{GROUP}} & \cellcolor{gray!6}{Conventional} & \cellcolor{gray!6}{5.77} & \cellcolor{gray!6}{1.00} & \cellcolor{gray!6}{109.0} & \cellcolor{gray!6}{0.10}\\
{INTUBATION}&{GROUP} & Non-Conventional & 15.2 & 2.37 & 302.8 & 0.02\\
{\cellcolor{gray!6}{INTUBATION}}&{\cellcolor{gray!6}{RADIOLOGICAL PATTERN}} & \cellcolor{gray!6}{Bronco. (Intercept)} & \cellcolor{gray!6}{0.95} & \cellcolor{gray!6}{0.61} & \cellcolor{gray!6}{1.48} & \cellcolor{gray!6}{0.82}\\
{INTUBATION}&{RADIOLOGICAL PATTERN} & Angio. & 0.73 & 0.37 & 1.40 & 0.34\\
{\cellcolor{gray!6}{INTUBATION}}&{\cellcolor{gray!6}{ICU}} & \cellcolor{gray!6}{No (Intercept)} & \cellcolor{gray!6}{0.00} & \cellcolor{gray!6}{0.00} & \cellcolor{gray!6}{Inf} & \cellcolor{gray!6}{0.99}\\
{INTUBATION}&{ICU} & Yes & 0.00 & 0.00 & Inf & 0.99\\
\midrule
{\cellcolor{gray!6}{ICU}}&{\cellcolor{gray!6}{GROUP}} & \cellcolor{gray!6}{Neutropenic (Intercept)} & \cellcolor{gray!6}{0.29} & \cellcolor{gray!6}{0.04} & \cellcolor{gray!6}{1.18} & \cellcolor{gray!6}{0.12}\\
{ICU}&{GROUP} & Conventional & 2.95 & 0.67 & 20.4 & 0.19\\
{\cellcolor{gray!6}{ICU}}&{\cellcolor{gray!6}{GROUP}} & \cellcolor{gray!6}{Non-Conventional} & \cellcolor{gray!6}{7.70} & \cellcolor{gray!6}{1.54} & \cellcolor{gray!6}{58.3} & \cellcolor{gray!6}{0.02}\\
{ICU}&{RADIOLOGICAL PATTERN} & Bronco. (Intercept) & 0.95 & 0.61 & 1.48 & 0.82\\
{\cellcolor{gray!6}{ICU}}&{\cellcolor{gray!6}{RADIOLOGICAL PATTERN}} & \cellcolor{gray!6}{Angio.} & \cellcolor{gray!6}{1.05} & \cellcolor{gray!6}{0.55} & \cellcolor{gray!6}{2.02} & \cellcolor{gray!6}{0.88}\\
\bottomrule
\end{tabular}
}
\end{table}

\label{sec:case1}
We start investigating the effect of the groups' allocation and the radiological pattern on the patient trajectory in the hospital: first, if they enter the ICU; second, if they are intubated; and, ultimately, whether they die.

Table \ref{table:logistic} reports the results of the univariate logistic regression analysis, suggesting that access to ICU (OR = 2.85) and intubation (OR = 2.63) are the only two significant predictors of death. Concerning intubation and access to ICU, patients of the non-conventional group have a much higher risk (OR INTUBATION = 15.2; OR ICU = 7.70), possibly because they tend to have longer diagnosis times since they are usually not considered at risk of AFF-IFI.

 Table \ref{table:lasso} reports the estimated ORs and, except for the interaction RP:ICU, all two-way interactions are estimated not to be relevant (OR = 1). Moreover, as a discriminative model, logistic regression cannot provide additional information about the relationships between risk predictors.

\begin{table}
\centering
  \caption{Estimated ORs for LASSO logistic regression with 2-way interactions to predict DEATH. Only ORs different from 1 are reported.\label{table:lasso}}
  \scalebox{0.8}{
\begin{tabular}{cc}
\toprule
Term & OR\\
\midrule
{{Intercept}} & {0.47}\\
%{GROUP = conventional} & 1.00\\
%{\cellcolor{gray!6}{GROUP = non-conventional}} & \cellcolor{gray!6}{1.00}\\
%{RADIOLOGICAL PATTERN = Angio.} & 1.00\\
{\cellcolor{gray!6}{ICU = yes}} & \cellcolor{gray!6}{1.65}\\
%{INTUBATION = yes} & 1.00\\
%{\cellcolor{gray!6}{GROUP = conventional \& RADIOLOGICAL PATTERN = Angio.}} & \cellcolor{gray!6}{1.00}\\
%{GROUP = non-conventional \& RADIOLOGICAL PATTERN = Angio.} & 1.00\\
%{\cellcolor{gray!6}{GROUP = conventional \& ICU = yes}} & \cellcolor{gray!6}{1.00}\\
%{GROUP = non-conventional \& ICU = yes} & 1.00\\
%{\cellcolor{gray!6}{GROUP = conventional \& INTUBATION = yes}} & \cellcolor{gray!6}{1.00}\\
%{GROUP = non-conventional \& INTUBATION = yes} & 1.00\\
{{RP = Angio. \& ICU = yes}} & {1.28}\\
%{RADIOLOGICAL PATTERN = Angio. \& INTUBATION = yes} & 1.00\\
%{\cellcolor{gray!6}{ICU = yes \& INTUBATION = yes}} & \cellcolor{gray!6}{1.00}\\
\bottomrule
\end{tabular}
}
\end{table}

\begin{figure}
\centering
\includegraphics[scale =0.8]{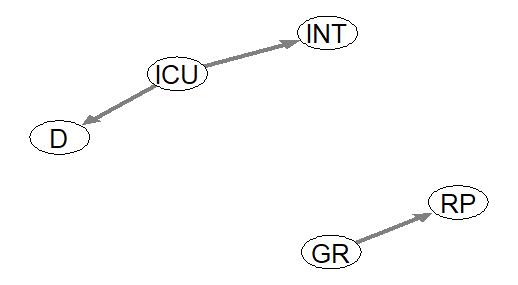}
 \caption{BN over the variables GROUP, ANGIOINVASIVE PATTERN, ICU, INTUBATION, and DEATH.\label{fig:bn1}}
\end{figure}

Figure~\ref{fig:bn1} reports the learned BN, which suggests that the group (GR) and the radiological pattern (RP) are independent of the patient's trajectory. Indeed, the group affects the radiological pattern and intubation is independent of death conditional on ICU. 
For completeness, we also try other classical DAG learning algorithms, such as the order stable version of the PC algorithm~\citep{colombo14a} and 
the max-min hill-climbing hybrid algorithm~\citep{tsamardinos2006max}, which both obtain the simpler DAG with the only edges GR$\rightarrow$RP and 
ICU$\rightarrow$INT.

\begin{figure}
\centering
\scalebox{0.43}{
\input{trees/tree1}
}
\caption{Staged tree over the variables GR, RP, ICU, INT, and DTH.
The colors of the nodes denote the 
stages and the labels of the edges indicate the corresponding events and in parenthesis the estimated probabilities.
\label{fig:lifehc}}
\end{figure}
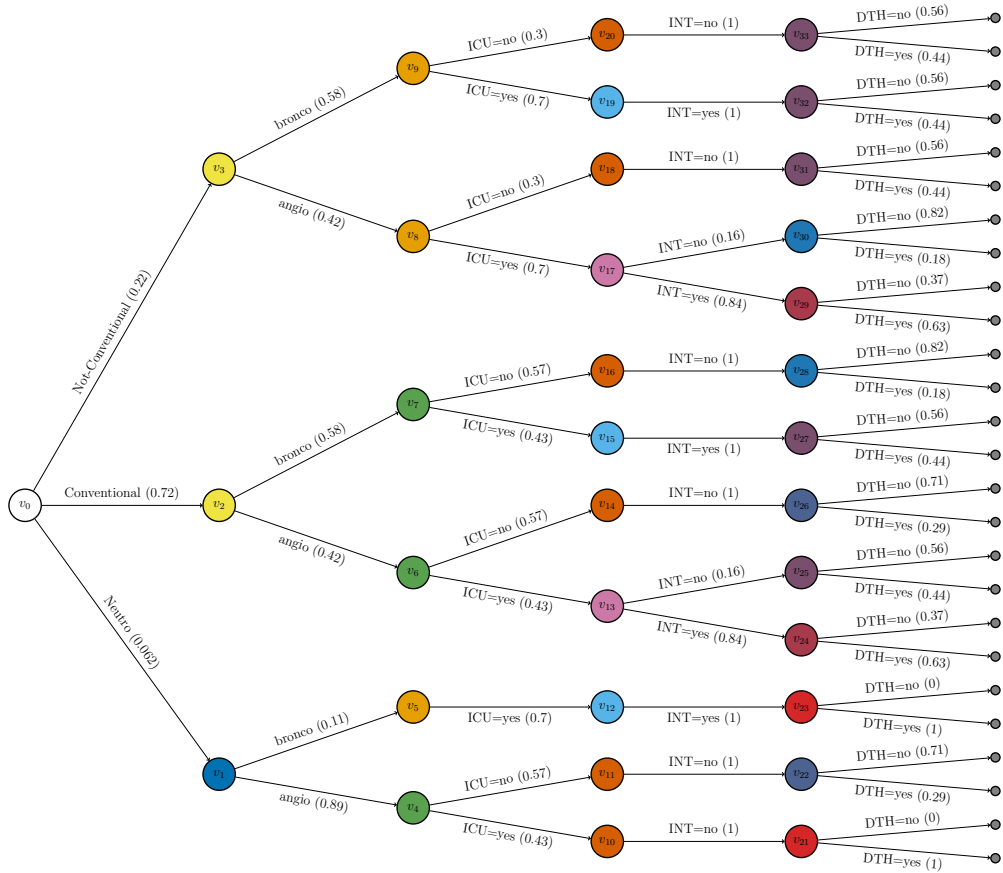

Figure \ref{fig:lifehc} reports the staged tree learned with a hill-climbing greedy algorithm minimizing the BIC  \citep{gorgen2022curved} using the \texttt{stagedtrees} R package \citep{Carli2022}.

The tree is highly expressive and intuitively shows:
\begin{itemize}
    \item The presence of combinations of variables not observed in the data. For instance, all patients in the neutropenic group with a broncoinvasive pattern enter the ICU.
    \item A non-product sample space. It is known that patients who do not enter the ICU cannot be intubated. For this reason, after the edge ICU = no, the edge INT = no 
    is always the only option.
    \item A highly non-symmetric pattern of dependence by the coloring of internal vertices, which BNs cannot generally represent.
\end{itemize}

Clinically we observe the following.
Patients in the non-conventional and conventional groups are equally likely to have the same radiological pattern ($v_2$-$v_3$). In particular, these two groups of patients are more likely to have a broncoinvasive pattern (58\%), whereas neutropenic patients are more likely to have an angioinvasive pattern (89\%). Admission to the ICU does not depend on the radiological pattern in the case of unconventional and conventional patients ($v_6$-$v_9$), but admission itself is higher in unconventional patients (70\%). On the other hand, the radiological pattern does affect the probability of admission to the ICU in the case of neutropenic patients. Neutropenic patients with a broncoinvasive pattern have a higher probability (70\%) of accessing the ICU than patients with an angioinvasive pattern (43\%). Thanks to the colors, we can see that the probability of ICU admission of neutropenic patients with a broncoinvasive pattern and non-conventional patients is the same. The same observation holds for neutropenic patients with an angioinvasive pattern, and conventional patients.

Upon admission to the ICU, patients with a broncoinvasive pattern have a 100\% chance of being intubated regardless of the group ($v_{10},v_{12},v_{14}$). Patients in the non-conventional and conventional groups are estimated to have the same chance of being intubated (82\%), while no neutropenic patients with an angioinvasive pattern are intubated.

Intubated patients with an angioinvasive pattern from the not-conventional and conventional groups have the same estimated  chance of dying, which reaches 63\% ($v_{24}$ and $v_{29}$). In contrast, the patients with the highest survival (82\%) are non-intubated individuals from the not-conventional group with an angioinvasive pattern and ICU admission ($v_{30}$) or conventional patients with a broncoinvasive pattern without ICU admission ($v_{28}$). Patients in the non-conventional group who have not been admitted to the ICU have a survival of 56\%, regardless of the radiological pattern ($v_{31}, v_{33}$). Not-conventional and conventional patients with a broncoinvasive pattern admitted to the ICU and intubated  ($v_{27},v_{32}$) have the same estimated survival probability of 56\%. Equally 56\% survival probability is found in not intubated ICU patients with an angioinvasive pattern in the conventional group ($v_{25}$). Conventional and neutropenic patients who have not been admitted to the ICU with an angioinvasive pattern have instead a 71\% probability of survival 
($v_{22}$, $v_{26}$). All neutropenic patients admitted to the ICU have died regardless of the radiological pattern and intubation 
($v_{21}$, $v_{23}$).

\subsection{2nd Case Study}

\label{sec:case2}
The previous case study is highly suited for staged tree modeling since it includes a limited number of variables, with an explicit causal ordering and an asymmetric sample space that the tree can explicitly and intuitively represent. However, recent advances in staged tree theory have made them a viable and efficient tool to investigate dependence in more complex scenarios, including a more extensive array of risk factors, as we showcase in the following data application. 

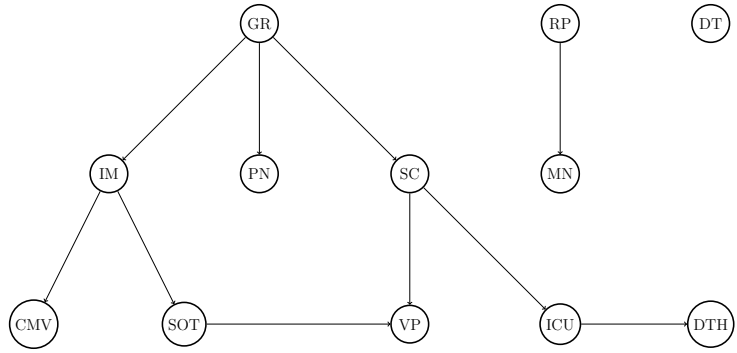
\begin{figure}
    \centering
    \scalebox{0.5}{
    \begin{tikzpicture}[auto, scale=20,
	node/.style={circle,inner sep=1mm,minimum size=1cm,draw,black,very thick,text=black}]

    \node [node] (GR) at (0,0) {GR};
    \node [node]  (RP) at (0.4,0) {RP};
    \node [node]  (DT) at (0.6,0) {DT};
    \node [node]  (IM) at (-0.2,-0.2) {IM};
    \node [node]  (PN) at (0,-0.2) {PN};
    \node [node]  (SC) at (0.2,-0.2) {SC};
    \node [node]  (MN) at (0.4,-0.2) {MN};
    \node [node]  (CMV) at (-0.3,-0.4) {CMV};
    \node [node]  (SOT) at (-0.1,-0.4) {SOT};
    \node [node]  (VP) at (0.2,-0.4) {VP};
    \node [node]  (ICU) at (0.4,-0.4) {ICU};
 \node [node]  (DTH) at (0.6,-0.4) {DTH};
   \draw[->] (GR) -- (IM);
   \draw[->] (GR) -- (PN);
   \draw[->] (GR) -- (SC);
   \draw[->] (RP) -- (MN);
    \draw[->] (IM) -- (CMV);
    \draw[->] (IM) -- (SOT);
    \draw[->] (SC) -- (ICU);
    \draw[->] (SC) -- (VP);
    \draw[->] (ICU) -- (DTH);
    \draw[->] (SOT) -- (VP);
    \end{tikzpicture}
    }
    \caption{Structure of the Bayesian network learned over AFF-IFI death-related risk factors.}
    \label{fig:bn2}
\end{figure}

We now consider all risk factors included in Table \ref{table:variables}, except for INT. Observations with missing values were dropped, giving 131 patients. Figure \ref{fig:bn2} reports the BN learned using the same procedure as in Section \ref{sec:case1}, where edges from DEATH, ICU, and DT to other variables are forbidden for ease of interpretation. ICU is the only parent of DEATH: thus, all other risk factors are independent of DEATH conditionally on ICU. Similarly, conditional on SYSTEMIC CORTICOIDS, all other risk factors, except ICU, are independent of DEATH. RADIOLOGICAL PATTERN, MALNUTRITION, and DIAGNOSTIC TIME are marginally independent of DEATH. This implies, for instance, that angioinvasive and broncoinvasive patients have the same probability of dying. The BN provides an intuitive representation of the variables' dependence and an efficient platform to answer inferential queries. For instance, we can straightforwardly compute the probability that a patient who enters the ICU has a solid organ transplant (probability equal to 0.35). Similarly, we can compute any other probability of interest from the model.

\begin{figure}
    \centering
        \scalebox{0.5}{
    \begin{tikzpicture}[auto, scale=20,
	node/.style={circle,inner sep=1mm,minimum size=1cm,draw,black,very thick,text=black}]

    \node [node] (GR) at (0,0) {GR};
    \node [node]  (RP) at (0.4,0) {RP};
    \node [node]  (DT) at (0.6,0) {DT};
    \node [node]  (IM) at (-0.2,-0.2) {IM};
    \node [node]  (PN) at (0,-0.2) {PN};
    \node [node]  (SC) at (0.2,-0.2) {SC};
    \node [node]  (MN) at (0.4,-0.2) {MN};
    \node [node]  (CMV) at (-0.3,-0.4) {CMV};
    \node [node]  (SOT) at (-0.1,-0.4) {SOT};
    \node [node]  (VP) at (0.2,-0.4) {VP};
    \node [node]  (ICU) at (0.4,-0.4) {ICU};
 \node [node]  (DTH) at (0.6,-0.4) {DTH};
   \draw[->] (GR) -- (IM);
   \draw[->] (GR) -- (PN);
   \draw[->] (GR) -- (SC);
   \path[->] (GR) edge [bend right = 30] (CMV);
   \draw[->] (GR) -- (RP);
   \path[->] (GR) edge [bend left = 30] (DT);
   \draw[->] (GR) -- (SOT);
   \draw[->] (MN) -- (ICU);
   \draw[->] (RP) -- (MN);
    \draw[->] (IM) -- (CMV);
    \path[->] (IM) edge [ bend right = 30] (MN);
    \draw[->] (SC) -- (ICU);
    \draw[->] (SC) -- (VP);
    \draw[->] (SC) -- (RP);
    \path[->] (SC) edge [ bend right = 20] (IM);
    \draw[->] (ICU) -- (DTH);
    \draw[->] (DT) -- (DTH);
    \draw[->] (SOT) -- (VP);
    \path[->] (VP) edge [bend right = 30] (DT);
    \end{tikzpicture}
    }
    \caption{Minimal DAG associated to learned staged tree over AFF-IFI death-related risk factors.}
    \label{fig:aldag}
\end{figure}
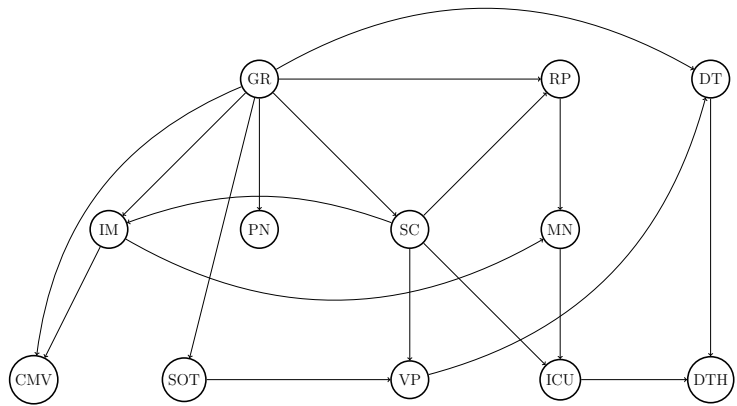

 Figure \ref{fig:aldag} reports the minimal DAG of the 2-parents staged tree over the AFF-IFI death-related risk learned using the algorithm of \citet{leonelli2023learning}. We chose two parents to find a balance between the goodness of fit and ease of interpretation. This minimal DAG reveals a much more involved dependence pattern. DIAGNOSTIC TIME, assumed to be marginally independent of DEATH by the BN, directly influences it. GROUP is a central variable that directly affects DIAGNOSTIC TIME, SYSTEMIC CORTICOIDS, RADIOLOGICAL PATTERN, IMMUNOTHERAPY, SOLID ORGAN TRANSPLANT, CMV INFECTION, and VIRAL PNEUMONIA. No variables are assumed to be marginally independent of DEATH. 

The 2-parents staged tree better fits the data, having a BIC of 1706.433, against that of the BN equal to 1758.461. Given the small dataset, checking predictive accuracy would be unreliable. However, staged trees have been shown to often outperform BNs in predictive tasks~\citep{carli2020new}.

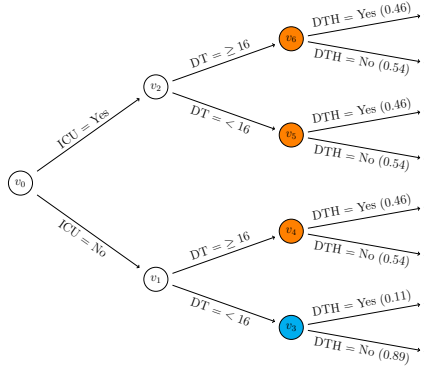
\begin{figure}
\centering
\scalebox{0.4}{
\begin{tikzpicture}
\renewcommand{\xx}{4.5}
\renewcommand{\yy}{0.8}
\node (v0) at (0*\xx,0*\yy) {\sage{white}{0}};
\node (v1) at (1*\xx,-4*\yy) {\sage{white}{1}};
\node (v2) at (1*\xx,4*\yy) {\sage{white}{2}};
\node (v3) at (2*\xx,-6*\yy) {\sage{cyan}{3}};
\node (v4) at (2*\xx,-2*\yy) {\sage{orange}{4}};
\node (v5) at (2*\xx,2*\yy) {\sage{orange}{5}};
\node (v6) at (2*\xx,6*\yy) {\sage{orange}{6}};
\node (v7) at (3*\xx,-7*\yy) {\leaf};
\node (v8) at (3*\xx,-5*\yy) {\leaf};
\node (v9) at (3*\xx,-3*\yy) {\leaf};
\node (v10) at (3*\xx,-1*\yy) {\leaf};
\node (v11) at (3*\xx,1*\yy) {\leaf};
\node (v12) at (3*\xx,3*\yy) {\leaf};
\node (v13) at (3*\xx,5*\yy) {\leaf};
\node (v14) at (3*\xx,7*\yy) {\leaf};
\draw[->] (v0) -- node [below, sloped] {{ICU = No }} (v1);
\draw[->] (v0) -- node [above, sloped] {{ICU = Yes}} (v2);
\draw[->] (v1) --  node [below, sloped] {{DT = $< 16$}} (v3);
\draw[->] (v1) --  node [above, sloped] {{DT = $\geq 16$}} (v4);
\draw[->] (v2) --  node [below, sloped] {{DT = $< 16$}} (v5);
\draw[->] (v2) --  node [above, sloped] {{DT = $\geq 16$}} (v6);
\draw[->] (v3) --  node [below, sloped] {{DTH = No (0.89)}} (v7);
\draw[->] (v3) --  node [above, sloped] {{DTH = Yes (0.11)}} (v8);
\draw[->] (v4) --  node [below, sloped] {{DTH = No (0.54)}} (v9);
\draw[->] (v4) --  node [above, sloped] {{DTH = Yes (0.46)}} (v10);
\draw[->] (v5) --  node [below, sloped] {{DTH = No (0.54)}} (v11);
\draw[->] (v5) --  node [above, sloped] {{DTH = Yes (0.46)}} (v12);
\draw[->] (v6) --  node [below, sloped] {{DTH = No (0.54)}} (v13);
\draw[->] (v6) --  node [above, sloped] {{DTH = Yes (0.46)}} (v14);
\end{tikzpicture}
}
\caption{Dependence subtree for DEATH from the minimal DAG of Figure \ref{fig:aldag}.\label{fig:depsub}}
\end{figure}

The minimal DAG of the staged tree provides a compressed, partial vision of the staged tree dependence structure. However, such an asymmetric structure is still learned from data. It can be visualized for each variable using a \textbf{dependence subtree} \citep{varando2021staged}, reporting the conditional independence coloring of a variable given its parents. Figure \ref{fig:depsub} reports the dependence subtree associated with the variable DEATH in the minimal DAG of Figure \ref{fig:aldag}. It shows that DEATH is conditionally independent of DIAGNOSTIC TIME given ICU = yes. Patients with a short diagnostic time who do not access the ICU have the smallest probability of death (11\%).

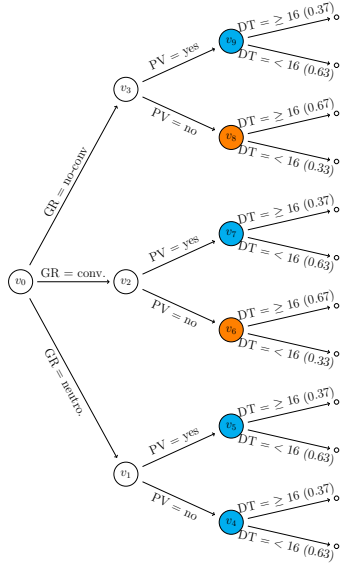
\begin{figure}
\centering
\scalebox{0.4}{
\begin{tikzpicture}
\renewcommand{\xx}{3.5}
\renewcommand{\yy}{0.8}
\node (v0) at (0*\xx,0*\yy) {\sage{white}{0}};
\node (v1) at (1*\xx,-8*\yy) {\sage{white}{1}};
\node (v2) at (1*\xx,0*\yy) {\sage{white}{2}};
\node (v3) at (1*\xx,8*\yy) {\sage{white}{3}};
\node (v4) at (2*\xx,-10*\yy) {\sage{cyan}{4}};
\node (v5) at (2*\xx,-6*\yy) {\sage{cyan}{5}};
\node (v6) at (2*\xx,-2*\yy) {\sage{orange}{6}};
\node (v7) at (2*\xx,2*\yy) {\sage{cyan}{7}};
\node (v8) at (2*\xx,6*\yy) {\sage{orange}{8}};
\node (v9) at (2*\xx,10*\yy) {\sage{cyan}{9}};
\node (v10) at (3*\xx,-11*\yy) {\leaf};
\node (v11) at (3*\xx,-9*\yy) {\leaf};
\node (v12) at (3*\xx,-7*\yy) {\leaf};
\node (v13) at (3*\xx,-5*\yy) {\leaf};
\node (v14) at (3*\xx,-3*\yy) {\leaf};
\node (v15) at (3*\xx,-1*\yy) {\leaf};
\node (v16) at (3*\xx,1*\yy) {\leaf};
\node (v17) at (3*\xx,3*\yy) {\leaf};
\node (v18) at (3*\xx,5*\yy) {\leaf};
\node (v19) at (3*\xx,7*\yy) {\leaf};
\node (v20) at (3*\xx,9*\yy) {\leaf};
\node (v21) at (3*\xx,11*\yy) {\leaf};
\draw[->] (v0) -- node [below, sloped] {{GR = neutro. }} (v1);
\draw[->] (v0) -- node [above, sloped] {{GR = conv.}} (v2);
\draw[->] (v0) -- node [above, sloped] {{GR = no-conv}} (v3);
\draw[->] (v1) --  node [below, sloped] {{PV = no}} (v4);
\draw[->] (v1) --  node [above, sloped] {{PV = yes}} (v5);
\draw[->] (v2) --  node [below, sloped] {{PV = no}} (v6);
\draw[->] (v2) --  node [above, sloped] {{PV = yes}} (v7);
\draw[->] (v3) --  node [below, sloped] {{PV = no }} (v8);
\draw[->] (v3) --  node [above, sloped] {{PV = yes }} (v9);
\draw[->] (v4) --  node [below, sloped] {{DT = $<$ 16 (0.63)}} (v10);
\draw[->] (v4) --  node [above, sloped] {{DT = $\geq$ 16 (0.37)}} (v11);
\draw[->] (v5) --  node [below, sloped] {{DT = $<$ 16 (0.63)}} (v12);
\draw[->] (v5) --  node [above, sloped] {{DT = $\geq$ 16 (0.37)}} (v13);
\draw[->] (v6) --  node [below, sloped] {{DT = $<$ 16 (0.33)}} (v14);
\draw[->] (v6) --  node [above, sloped] {{DT = $\geq$ 16 (0.67)}} (v15);
\draw[->] (v7) --  node [below, sloped] {{DT = $<$ 16 (0.63)}} (v16);
\draw[->] (v7) --  node [above, sloped] {{DT = $\geq$ 16 (0.37)}} (v17);
\draw[->] (v8) --  node [below, sloped] {{DT = $<$ 16 (0.33)}} (v18);
\draw[->] (v8) --  node [above, sloped] {{DT = $\geq$ 16 (0.67)}} (v19);
\draw[->] (v9) --  node [below, sloped] {{DT = $<$ 16 (0.63)}} (v20);
\draw[->] (v9) --  node [above, sloped] {{DT = $\geq$ 16 (0.37)}} (v21);
\end{tikzpicture}
}
\caption{Dependence subtree for DIAGNOSTIC TIME from the minimal DAG of Figure \ref{fig:aldag}.\label{fig:hola}}
\end{figure}

The group patients belong to, which has already been observed to be associated with the radiological pattern, has a direct influence on the diagnostic time, which in turn affects mortality. The dependence subtree for DIAGNOSTIC TIME in Figure \ref{fig:hola} shows that non-conventional patients, usually not considered by diagnostic criteria, tend to have a longer diagnostic time. For this reason, the use of new diagnostic criteria that consider the broncoinvasive radiological pattern and a broader classification of risk groups would lead to a rapid diagnosis of AFF-IFI patients, possibly entailing a reduction in mortality. In turn, the use of these criteria would lead to a reduction in hospital pressure in ICU (as observed in case study 1) with a consequent reduction of hospitality costs and an increase in survival \citep{menzin2009mortality}.

%Because staged trees model the entire distribution of the risk factors, they can provide various model-based measures of association and information. Table \ref{table:mi} reports the marginal mutual information between DEATH and each risk factor. As expected, ICU and DT are the risk factors with the strongest influence on death. Notice that in the case of the BN in Figure \ref{fig:bn2}, the mutual information between DT and DEATH would have been estimated as zero, since they are marginally independent. Table \ref{table:mi} further reports the mutual information between DEATH and each risk factor conditional on one risk factor being observed. In parenthesis, the variation from the marginal value is reported. For instance, GROUP has a mutual information of 0.045/100 with DEATH, which increases by 19.3\%, knowing whether a patient entered the ICU. Conversely, it decreases by 24.6\% conditional on the outcome of IMMUNOTHERAPY. Although standard for BNs, this mutual information analysis was never carried out for staged trees.

\section{Discussion}

Observations from the results on the first case study (Section~\ref{sec:case1}) reinforce what has been described in the literature, that non-neutropenic patients (conventional and not-conventional) tend to develop respiratory forms more frequently with bronchopulmonary patterns  \citep{liu2020airway,nucci2010probable,nucci2013early,park2010clinical}, although they differ considerably in the type of underlying diseases and degree of immunosuppression. On the other hand, neutropenic patients more frequently present angioinvasive patterns \citep{de2008revised,latge2019aspergillus}. Regarding actionable conclusions, the observations drawn from the staged tree highlight the need to include information about broncoinvasive patterns in the diagnostic criteria for AFF-IFI since the gold standard EORTC criteria \citep{de2008revised,donnelly2020revision} completely overlook them.  We are currently working on proposing extended diagnostic criteria based on the results of this study. However, their discussion is beyond the scope of this paper.

\section{Conclusions}

Recently staged trees have been the focus of research, leading to a better understanding of their underlying dependence structure, more flexible visualization platforms, more efficient data learning algorithms, and open-source implementations. Given all these advances, staged trees can now provide unique insights into data-driven health applications to support practitioners in the real world.

Two case studies in AFF-IFIs demonstrated the flexibility of staged trees in intuitively representing highly asymmetric patterns of dependence, which BNs cannot explicitly visualize. Furthermore, they provide an efficient platform to answer inferential and independence queries and sensitivity analyses. Using staged trees helped clinicians understand the relationship between risk factors in AFF-IFI more intuitively than BNs.

From a clinical perspective, the analysis showed that the non-conventional group, usually not considered by standard diagnostic scales, shares many characteristics and risks with the neutropenic and conventional groups. Therefore this observation highlights the need to construct more widely applicable entry criteria for diagnostic scales in AFF-IFIs for the timely diagnosis of this group of often overlooked patients. Furthermore, the analysis has shown that the broncoinvasive radiological pattern, not considered within the gold standard diagnostic criteria, plays a critical role in AFF-IFI. Its full clinical appraisal would lead to a timely diagnosis and, consequently, a decrease in mortality.

Despite the small sample size and the lack of statistical validation techniques, all the insights given by staged trees match the clinical intuition of the doctors
in our study and information from established literature \citep{liu2020airway,park2010clinical}. However, the staged tree provided a more intuitive platform for their interpretation and discussion among clinicians than numerical tables commonly reported in medical studies.

%% The Appendices part is started with the command \appendix;
%% appendix sections are then done as normal sections
%% \appendix

%% \section{}
%% \label{}

%% If you have bibdatabase file and want bibtex to generate the
%% bibitems, please use
%%
\bibliographystyle{elsarticle-harv} 
\bibliography{biblio}
\end{document}

%% file: preamble.tex
\usepackage[table, x11names]{xcolor}
\usepackage{pgf, tikz} % per grafici direttamente con latex
\usetikzlibrary{arrows,automata,fit}
\usetikzlibrary{shapes}
\usepackage{bm}
\usepackage{booktabs}

\newcommand{\xx}{1}
\newcommand{\yy}{1}
\newcommand{\sage}[2]{\tikz{\node[shape=circle,draw,inner sep=1pt,minimum width = 0.8cm, fill=#1]{$v_{#2}$};}}

\newcommand{\leaf}{\tikz{\node[shape=circle,draw,inner sep=1.5pt,fill=white] {};}}

\newcommand\independent{\protect\mathpalette{\protect\independenT}{\perp}}
\def\independenT#1#2{\mathrel{\rlap{$#1#2$}\mkern2mu{#1#2}}}

%% file: trees/tree_example.tex
\begin{tikzpicture}[auto, scale=12,
	A_NA/.style={circle,inner sep=1mm,minimum size=1cm,draw,black,very thick,fill={rgb,255:red,255; green,255; blue,255},text=black},
	B_1/.style={circle,inner sep=1mm,minimum size=1cm,draw,black,very thick,fill={rgb,255:red,78; green,121; blue,167},text=black},
	C_2/.style={circle,inner sep=1mm,minimum size=1cm,draw,black,very thick,fill={rgb,255:red,242; green,142; blue,43},text=black},
	C_3/.style={circle,inner sep=1mm,minimum size=1cm,draw,black,very thick,fill={rgb,255:red,225; green,87; blue,89},text=black},
	D_4/.style={circle,inner sep=1mm,minimum size=1cm,draw,black,very thick,fill={rgb,255:red,118; green,183; blue,178},text=black},
	D_5/.style={circle,inner sep=1mm,minimum size=1cm,draw,black,very thick,fill={rgb,255:red,89; green,161; blue,79},text=black},
	D_6/.style={circle,inner sep=1mm,minimum size=1cm,draw,black,very thick,fill={rgb,255:red,237; green,201; blue,72},text=black},
	 leaf/.style={circle,inner sep=1mm,minimum size=0.2cm,draw,very thick,black,fill=gray,text=black}]

	\node [A_NA] (root) at (0.000000, 0.500000)	{$v_{0}$};
	\node [B_1] (root-1) at (0.250000, 0.233333)	{$v_{1}$};
	\node [B_1] (root-0) at (0.250000, 0.766667)	{$v_{2}$};
	\node [C_2] (root-1-1) at (0.500000, 0.100000)	{$v_{3}$};
	\node [C_3] (root-1-0) at (0.500000, 0.366667)	{$v_{4}$};
	\node [C_3] (root-0-1) at (0.500000, 0.633333)	{$v_{5}$};
	\node [C_2] (root-0-0) at (0.500000, 0.900000)	{$v_{6}$};
	\node [D_4] (root-1-1-1) at (0.750000, 0.033333)	{$v_{7}$};
	\node [D_4] (root-1-1-0) at (0.750000, 0.166667)	{$v_{8}$};
	\node [D_4] (root-1-0-1) at (0.750000, 0.300000)	{$v_{9}$};
	\node [D_4] (root-1-0-0) at (0.750000, 0.433333)	{$v_{10}$};
	\node [D_5] (root-0-1-1) at (0.750000, 0.566667)	{$v_{11}$};
	\node [D_6] (root-0-1-0) at (0.750000, 0.700000)	{$v_{12}$};
	\node [D_5] (root-0-0-1) at (0.750000, 0.833333)	{$v_{13}$};
	\node [D_6] (root-0-0-0) at (0.750000, 0.966667)	{$v_{14}$};
	\node [leaf] (root-1-1-1-1) at (1.000000, 0.000000)	{};
	\node [leaf] (root-1-1-1-0) at (1.000000, 0.066667)	{};
	\node [leaf] (root-1-1-0-1) at (1.000000, 0.133333)	{};
	\node [leaf] (root-1-1-0-0) at (1.000000, 0.200000)	{};
	\node [leaf] (root-1-0-1-1) at (1.000000, 0.266667)	{};
	\node [leaf] (root-1-0-1-0) at (1.000000, 0.333333)	{};
	\node [leaf] (root-1-0-0-1) at (1.000000, 0.400000)	{};
	\node [leaf] (root-1-0-0-0) at (1.000000, 0.466667)	{};
	\node [leaf] (root-0-1-1-1) at (1.000000, 0.533333)	{};
	\node [leaf] (root-0-1-1-0) at (1.000000, 0.600000)	{};
	\node [leaf] (root-0-1-0-1) at (1.000000, 0.666667)	{};
	\node [leaf] (root-0-1-0-0) at (1.000000, 0.733333)	{};
	\node [leaf] (root-0-0-1-1) at (1.000000, 0.800000)	{};
	\node [leaf] (root-0-0-1-0) at (1.000000, 0.866667)	{};
	\node [leaf] (root-0-0-0-1) at (1.000000, 0.933333)	{};
	\node [leaf] (root-0-0-0-0) at (1.000000, 1.000000)	{};

	\draw[->] (root) -- node [sloped,swap]{A=1} (root-1);
	\draw[->] (root) -- node [sloped]{A=0} (root-0);
	\draw[->] (root-1) -- node [sloped,swap]{B=1} (root-1-1);
	\draw[->] (root-1) -- node [sloped]{B=0} (root-1-0);
	\draw[->] (root-0) -- node [sloped,swap]{B=1} (root-0-1);
	\draw[->] (root-0) -- node [sloped]{B=0} (root-0-0);
	\draw[->] (root-1-1) -- node [sloped,swap]{C=1} (root-1-1-1);
	\draw[->] (root-1-1) -- node [sloped]{C=0} (root-1-1-0);
	\draw[->] (root-1-0) -- node [sloped,swap]{C=1} (root-1-0-1);
	\draw[->] (root-1-0) -- node [sloped]{C=0} (root-1-0-0);
	\draw[->] (root-0-1) -- node [sloped,swap]{C=1} (root-0-1-1);
	\draw[->] (root-0-1) -- node [sloped]{C=0} (root-0-1-0);
	\draw[->] (root-0-0) -- node [sloped,swap]{C=1} (root-0-0-1);
	\draw[->] (root-0-0) -- node [sloped]{C=0} (root-0-0-0);
	\draw[->] (root-1-1-1) -- node [sloped,swap]{D=1} (root-1-1-1-1);
	\draw[->] (root-1-1-1) -- node [sloped]{D=0} (root-1-1-1-0);
	\draw[->] (root-1-1-0) -- node [sloped,swap]{D=1} (root-1-1-0-1);
	\draw[->] (root-1-1-0) -- node [sloped]{D=0} (root-1-1-0-0);
	\draw[->] (root-1-0-1) -- node [sloped,swap]{D=1} (root-1-0-1-1);
	\draw[->] (root-1-0-1) -- node [sloped]{D=0} (root-1-0-1-0);
	\draw[->] (root-1-0-0) -- node [sloped,swap]{D=1} (root-1-0-0-1);
	\draw[->] (root-1-0-0) -- node [sloped]{D=0} (root-1-0-0-0);
	\draw[->] (root-0-1-1) -- node [sloped,swap]{D=1} (root-0-1-1-1);
	\draw[->] (root-0-1-1) -- node [sloped]{D=0} (root-0-1-1-0);
	\draw[->] (root-0-1-0) -- node [sloped,swap]{D=1} (root-0-1-0-1);
	\draw[->] (root-0-1-0) -- node [sloped]{D=0} (root-0-1-0-0);
	\draw[->] (root-0-0-1) -- node [sloped,swap]{D=1} (root-0-0-1-1);
	\draw[->] (root-0-0-1) -- node [sloped]{D=0} (root-0-0-1-0);
	\draw[->] (root-0-0-0) -- node [sloped,swap]{D=1} (root-0-0-0-1);
	\draw[->] (root-0-0-0) -- node [sloped]{D=0} (root-0-0-0-0);
\end{tikzpicture}

%% file: trees/tree1.tex
\begin{tikzpicture}[auto, scale=20,
	GR_NA/.style={circle,inner sep=1mm,minimum size=1cm,draw,black,very thick,fill={rgb,255:red,255; green,255; blue,255},text=black},
	RP_1/.style={circle,inner sep=1mm,minimum size=1cm,draw,black,very thick,fill={rgb,255:red,0; green,114; blue,178},text=black},
	RP_2/.style={circle,inner sep=1mm,minimum size=1cm,draw,black,very thick,fill={rgb,255:red,240; green,228; blue,66},text=black},
	ICU_3/.style={circle,inner sep=1mm,minimum size=1cm,draw,black,very thick,fill={rgb,255:red,89; green,161; blue,79},text=black},
	ICU_4/.style={circle,inner sep=1mm,minimum size=1cm,draw,black,very thick,fill={rgb,255:red,230; green,159; blue,0},text=black},
	INT_5/.style={circle,inner sep=1mm,minimum size=1cm,draw,black,very thick,fill={rgb,255:red,213; green,94; blue,0},text=black},
	INT_6/.style={circle,inner sep=1mm,minimum size=1cm,draw,black,very thick,fill={rgb,255:red,86; green,180; blue,233},text=black},
	INT_UNOBSERVED/.style={circle,inner sep=1mm,minimum size=1cm,draw,black,very thick,fill={rgb,255:red,255; green,255; blue,255},text=black},
	INT_7/.style={circle,inner sep=1mm,minimum size=1cm,draw,black,very thick,fill={rgb,255:red,204; green,121; blue,167},text=black},
	D_UNOBSERVED/.style={circle,inner sep=1mm,minimum size=1cm,draw,black,very thick,fill={rgb,255:red,255; green,255; blue,255},text=black},
	D_8/.style={circle,inner sep=1mm,minimum size=1cm,draw,black,very thick,fill={rgb,255:red,214; green,39; blue,40},text=black},
	D_9/.style={circle,inner sep=1mm,minimum size=1cm,draw,black,very thick,fill={rgb,255:red,76; green,99; blue,145},text=black},
	D_10/.style={circle,inner sep=1mm,minimum size=1cm,draw,black,very thick,fill={rgb,255:red,168; green,59; blue,75},text=black},
	D_11/.style={circle,inner sep=1mm,minimum size=1cm,draw,black,very thick,fill={rgb,255:red,122; green,79; blue,110},text=black},
	D_12/.style={circle,inner sep=1mm,minimum size=1cm,draw,black,very thick,fill={rgb,255:red,31; green,119; blue,180},text=black},
	 leaf/.style={circle,inner sep=1mm,minimum size=0.2cm,draw,very thick,black,fill=gray,text=black}]

	\node [GR_NA] (root) at (0.000000, 0.546000)	{$v_{0}$};
	\node [RP_1] (root-Neutro) at (0.300000, 0.130000)	{$v_{1}$};
	\node [RP_2] (root-Conventional) at (0.300000, 0.546000)	{$v_{2}$};
	\node [RP_2] (root-Not-Conventional) at (0.300000, 1.066000)	{$v_{3}$};
	\node [ICU_3] (root-Neutro-angio) at (0.600000, 0.078000)	{$v_{4}$};
	\node [ICU_4] (root-Neutro-bronco) at (0.600000, 0.234000)	{$v_{5}$};
	\node [ICU_3] (root-Conventional-angio) at (0.600000, 0.442000)	{$v_{6}$};
	\node [ICU_3] (root-Conventional-bronco) at (0.600000, 0.702000)	{$v_{7}$};
	\node [ICU_4] (root-Not-Conventional-angio) at (0.600000, 0.962000)	{$v_{8}$};
	\node [ICU_4] (root-Not-Conventional-bronco) at (0.600000, 1.222000)	{$v_{9}$};
	\node [INT_5] (root-Neutro-angio-yes) at (0.900000, 0.026000)	{$v_{10}$};
	\node [INT_5] (root-Neutro-angio-no) at (0.900000, 0.130000)	{$v_{11}$};
	\node [INT_6] (root-Neutro-bronco-yes) at (0.900000, 0.234000)	{$v_{12}$};
	\node [INT_7] (root-Conventional-angio-yes) at (0.900000, 0.390000)	{$v_{13}$};
	\node [INT_5] (root-Conventional-angio-no) at (0.900000, 0.546000)	{$v_{14}$};
	\node [INT_6] (root-Conventional-bronco-yes) at (0.900000, 0.650000)	{$v_{15}$};
	\node [INT_5] (root-Conventional-bronco-no) at (0.900000, 0.754000)	{$v_{16}$};
	\node [INT_7] (root-Not-Conventional-angio-yes) at (0.900000, 0.910000)	{$v_{17}$};
	\node [INT_5] (root-Not-Conventional-angio-no) at (0.900000, 1.066000)	{$v_{18}$};
	\node [INT_6] (root-Not-Conventional-bronco-yes) at (0.900000, 1.170000)	{$v_{19}$};
	\node [INT_5] (root-Not-Conventional-bronco-no) at (0.900000, 1.274000)	{$v_{20}$};
	\node [D_8] (root-Neutro-angio-yes-no) at (1.200000, 0.026000)	{$v_{21}$};
	\node [D_9] (root-Neutro-angio-no-no) at (1.200000, 0.130000)	{$v_{22}$};
	\node [D_8] (root-Neutro-bronco-yes-yes) at (1.200000, 0.234000)	{$v_{23}$};
	\node [D_10] (root-Conventional-angio-yes-yes) at (1.200000, 0.338000)	{$v_{24}$};
	\node [D_11] (root-Conventional-angio-yes-no) at (1.200000, 0.442000)	{$v_{25}$};
	\node [D_9] (root-Conventional-angio-no-no) at (1.200000, 0.546000)	{$v_{26}$};
	\node [D_11] (root-Conventional-bronco-yes-yes) at (1.200000, 0.650000)	{$v_{27}$};
	\node [D_12] (root-Conventional-bronco-no-no) at (1.200000, 0.754000)	{$v_{28}$};
	\node [D_10] (root-Not-Conventional-angio-yes-yes) at (1.200000, 0.858000)	{$v_{29}$};
	\node [D_12] (root-Not-Conventional-angio-yes-no) at (1.200000, 0.962000)	{$v_{30}$};
	\node [D_11] (root-Not-Conventional-angio-no-no) at (1.200000, 1.066000)	{$v_{31}$};
	\node [D_11] (root-Not-Conventional-bronco-yes-yes) at (1.200000, 1.170000)	{$v_{32}$};
	\node [D_11] (root-Not-Conventional-bronco-no-no) at (1.200000, 1.274000)	{$v_{33}$};
	\node [leaf] (root-Neutro-angio-yes-no-yes) at (1.500000, 0.000000)	{};
	\node [leaf] (root-Neutro-angio-yes-no-no) at (1.500000, 0.052000)	{};
	\node [leaf] (root-Neutro-angio-no-no-yes) at (1.500000, 0.104000)	{};
	\node [leaf] (root-Neutro-angio-no-no-no) at (1.500000, 0.156000)	{};
	\node [leaf] (root-Neutro-bronco-yes-yes-yes) at (1.500000, 0.208000)	{};
	\node [leaf] (root-Neutro-bronco-yes-yes-no) at (1.500000, 0.260000)	{};
	\node [leaf] (root-Conventional-angio-yes-yes-yes) at (1.500000, 0.312000)	{};
	\node [leaf] (root-Conventional-angio-yes-yes-no) at (1.500000, 0.364000)	{};
	\node [leaf] (root-Conventional-angio-yes-no-yes) at (1.500000, 0.416000)	{};
	\node [leaf] (root-Conventional-angio-yes-no-no) at (1.500000, 0.468000)	{};
	\node [leaf] (root-Conventional-angio-no-no-yes) at (1.500000, 0.520000)	{};
	\node [leaf] (root-Conventional-angio-no-no-no) at (1.500000, 0.572000)	{};
	\node [leaf] (root-Conventional-bronco-yes-yes-yes) at (1.500000, 0.624000)	{};
	\node [leaf] (root-Conventional-bronco-yes-yes-no) at (1.500000, 0.676000)	{};
	\node [leaf] (root-Conventional-bronco-no-no-yes) at (1.500000, 0.728000)	{};
	\node [leaf] (root-Conventional-bronco-no-no-no) at (1.500000, 0.780000)	{};
	\node [leaf] (root-Not-Conventional-angio-yes-yes-yes) at (1.500000, 0.832000)	{};
	\node [leaf] (root-Not-Conventional-angio-yes-yes-no) at (1.500000, 0.884000)	{};
	\node [leaf] (root-Not-Conventional-angio-yes-no-yes) at (1.500000, 0.936000)	{};
	\node [leaf] (root-Not-Conventional-angio-yes-no-no) at (1.500000, 0.988000)	{};
	\node [leaf] (root-Not-Conventional-angio-no-no-yes) at (1.500000, 1.040000)	{};
	\node [leaf] (root-Not-Conventional-angio-no-no-no) at (1.500000, 1.092000)	{};
	\node [leaf] (root-Not-Conventional-bronco-yes-yes-yes) at (1.500000, 1.144000)	{};
	\node [leaf] (root-Not-Conventional-bronco-yes-yes-no) at (1.500000, 1.196000)	{};
	\node [leaf] (root-Not-Conventional-bronco-no-no-yes) at (1.500000, 1.248000)	{};
	\node [leaf] (root-Not-Conventional-bronco-no-no-no) at (1.500000, 1.300000)	{};

	\draw[->] (root) -- node [sloped]{Neutro (0.062)} (root-Neutro);
	\draw[->] (root) -- node [sloped]{Conventional (0.72)} (root-Conventional);
	\draw[->] (root) -- node [sloped]{Not-Conventional (0.22)} (root-Not-Conventional);
	\draw[->] (root-Neutro) -- node [sloped,swap]{angio (0.89)} (root-Neutro-angio);
	\draw[->] (root-Neutro) -- node [sloped]{bronco (0.11)} (root-Neutro-bronco);
	\draw[->] (root-Conventional) -- node [sloped,swap]{angio (0.42)} (root-Conventional-angio);
	\draw[->] (root-Conventional) -- node [sloped]{bronco (0.58)} (root-Conventional-bronco);
	\draw[->] (root-Not-Conventional) -- node [sloped,swap]{angio (0.42)} (root-Not-Conventional-angio);
	\draw[->] (root-Not-Conventional) -- node [sloped]{bronco (0.58)} (root-Not-Conventional-bronco);
	\draw[->] (root-Neutro-angio) -- node [sloped,swap]{ICU=yes (0.43)} (root-Neutro-angio-yes);
	\draw[->] (root-Neutro-angio) -- node [sloped]{ICU=no (0.57)} (root-Neutro-angio-no);
	\draw[->] (root-Neutro-bronco) -- node [sloped,swap]{ICU=yes (0.7)} (root-Neutro-bronco-yes);
	\draw[->] (root-Conventional-angio) -- node [sloped,swap]{ICU=yes (0.43)} (root-Conventional-angio-yes);
	\draw[->] (root-Conventional-angio) -- node [sloped]{ICU=no (0.57)} (root-Conventional-angio-no);
	\draw[->] (root-Conventional-bronco) -- node [sloped,swap]{ICU=yes (0.43)} (root-Conventional-bronco-yes);
	\draw[->] (root-Conventional-bronco) -- node [sloped]{ICU=no (0.57)} (root-Conventional-bronco-no);
	\draw[->] (root-Not-Conventional-angio) -- node [sloped,swap]{ICU=yes (0.7)} (root-Not-Conventional-angio-yes);
	\draw[->] (root-Not-Conventional-angio) -- node [sloped]{ICU=no (0.3)} (root-Not-Conventional-angio-no);
	\draw[->] (root-Not-Conventional-bronco) -- node [sloped,swap]{ICU=yes (0.7)} (root-Not-Conventional-bronco-yes);
	\draw[->] (root-Not-Conventional-bronco) -- node [sloped]{ICU=no (0.3)} (root-Not-Conventional-bronco-no);
	\draw[->] (root-Neutro-angio-yes) -- node [sloped]{INT=no (1)} (root-Neutro-angio-yes-no);
	\draw[->] (root-Neutro-angio-no) -- node [sloped]{INT=no (1)} (root-Neutro-angio-no-no);
	\draw[->] (root-Neutro-bronco-yes) -- node [sloped,swap]{INT=yes (1)} (root-Neutro-bronco-yes-yes);
	\draw[->] (root-Conventional-angio-yes) -- node [sloped,swap]{INT=yes (0.84)} (root-Conventional-angio-yes-yes);
	\draw[->] (root-Conventional-angio-yes) -- node [sloped]{INT=no (0.16)} (root-Conventional-angio-yes-no);
	\draw[->] (root-Conventional-angio-no) -- node [sloped]{INT=no (1)} (root-Conventional-angio-no-no);
	\draw[->] (root-Conventional-bronco-yes) -- node [sloped,swap]{INT=yes (1)} (root-Conventional-bronco-yes-yes);
	\draw[->] (root-Conventional-bronco-no) -- node [sloped]{INT=no (1)} (root-Conventional-bronco-no-no);
	\draw[->] (root-Not-Conventional-angio-yes) -- node [sloped,swap]{INT=yes (0.84)} (root-Not-Conventional-angio-yes-yes);
	\draw[->] (root-Not-Conventional-angio-yes) -- node [sloped]{INT=no (0.16)} (root-Not-Conventional-angio-yes-no);
	\draw[->] (root-Not-Conventional-angio-no) -- node [sloped]{INT=no (1)} (root-Not-Conventional-angio-no-no);
	\draw[->] (root-Not-Conventional-bronco-yes) -- node [sloped,swap]{INT=yes (1)} (root-Not-Conventional-bronco-yes-yes);
	\draw[->] (root-Not-Conventional-bronco-no) -- node [sloped]{INT=no (1)} (root-Not-Conventional-bronco-no-no);
	\draw[->] (root-Neutro-angio-yes-no) -- node [sloped,swap]{DTH=yes (1)} (root-Neutro-angio-yes-no-yes);
	\draw[->] (root-Neutro-angio-yes-no) -- node [sloped]{DTH=no (0)} (root-Neutro-angio-yes-no-no);
	\draw[->] (root-Neutro-angio-no-no) -- node [sloped,swap]{DTH=yes (0.29)} (root-Neutro-angio-no-no-yes);
	\draw[->] (root-Neutro-angio-no-no) -- node [sloped]{DTH=no (0.71)} (root-Neutro-angio-no-no-no);
	\draw[->] (root-Neutro-bronco-yes-yes) -- node [sloped,swap]{DTH=yes (1)} (root-Neutro-bronco-yes-yes-yes);
	\draw[->] (root-Neutro-bronco-yes-yes) -- node [sloped]{DTH=no (0)} (root-Neutro-bronco-yes-yes-no);
	\draw[->] (root-Conventional-angio-yes-yes) -- node [sloped,swap]{DTH=yes (0.63)} (root-Conventional-angio-yes-yes-yes);
	\draw[->] (root-Conventional-angio-yes-yes) -- node [sloped]{DTH=no (0.37)} (root-Conventional-angio-yes-yes-no);
	\draw[->] (root-Conventional-angio-yes-no) -- node [sloped,swap]{DTH=yes (0.44)} (root-Conventional-angio-yes-no-yes);
	\draw[->] (root-Conventional-angio-yes-no) -- node [sloped]{DTH=no (0.56)} (root-Conventional-angio-yes-no-no);
	\draw[->] (root-Conventional-angio-no-no) -- node [sloped,swap]{DTH=yes (0.29)} (root-Conventional-angio-no-no-yes);
	\draw[->] (root-Conventional-angio-no-no) -- node [sloped]{DTH=no (0.71)} (root-Conventional-angio-no-no-no);
	\draw[->] (root-Conventional-bronco-yes-yes) -- node [sloped,swap]{DTH=yes (0.44)} (root-Conventional-bronco-yes-yes-yes);
	\draw[->] (root-Conventional-bronco-yes-yes) -- node [sloped]{DTH=no (0.56)} (root-Conventional-bronco-yes-yes-no);
	\draw[->] (root-Conventional-bronco-no-no) -- node [sloped,swap]{DTH=yes (0.18)} (root-Conventional-bronco-no-no-yes);
	\draw[->] (root-Conventional-bronco-no-no) -- node [sloped]{DTH=no (0.82)} (root-Conventional-bronco-no-no-no);
	\draw[->] (root-Not-Conventional-angio-yes-yes) -- node [sloped,swap]{DTH=yes (0.63)} (root-Not-Conventional-angio-yes-yes-yes);
	\draw[->] (root-Not-Conventional-angio-yes-yes) -- node [sloped]{DTH=no (0.37)} (root-Not-Conventional-angio-yes-yes-no);
	\draw[->] (root-Not-Conventional-angio-yes-no) -- node [sloped,swap]{DTH=yes (0.18)} (root-Not-Conventional-angio-yes-no-yes);
	\draw[->] (root-Not-Conventional-angio-yes-no) -- node [sloped]{DTH=no (0.82)} (root-Not-Conventional-angio-yes-no-no);
	\draw[->] (root-Not-Conventional-angio-no-no) -- node [sloped,swap]{DTH=yes (0.44)} (root-Not-Conventional-angio-no-no-yes);
	\draw[->] (root-Not-Conventional-angio-no-no) -- node [sloped]{DTH=no (0.56)} (root-Not-Conventional-angio-no-no-no);
	\draw[->] (root-Not-Conventional-bronco-yes-yes) -- node [sloped,swap]{DTH=yes (0.44)} (root-Not-Conventional-bronco-yes-yes-yes);
	\draw[->] (root-Not-Conventional-bronco-yes-yes) -- node [sloped]{DTH=no (0.56)} (root-Not-Conventional-bronco-yes-yes-no);
	\draw[->] (root-Not-Conventional-bronco-no-no) -- node [sloped,swap]{DTH=yes (0.44)} (root-Not-Conventional-bronco-no-no-yes);
	\draw[->] (root-Not-Conventional-bronco-no-no) -- node [sloped]{DTH=no (0.56)} (root-Not-Conventional-bronco-no-no-no);
\end{tikzpicture}